\documentclass[aps,prb,twocolumn,showpacs,superscriptaddress]{revtex4-2}  
\usepackage{float}
\usepackage{bbm}
\usepackage{epsfig}
\usepackage{epstopdf}
\usepackage{graphicx}
\usepackage{amsmath,amssymb}
\usepackage{amsmath,bm}
\usepackage{physics}
\usepackage{color}
\usepackage{hyperref}
\usepackage{lineno,blindtext}
\usepackage{siunitx, booktabs}
\usepackage{diagbox, eqparbox, hhline}
\usepackage{soul}
\usepackage{xcolor}
\usepackage[normalem]{ulem}
\newcolumntype{P}[1]{>{\centering\arraybackslash}p{#1}}
\begin{document}

\title{Multitude of exceptional points in van der Waals magnets}

\author{Xin Li}
\email{licqp@bc.edu}
\author{Kuangyin Deng}
\author{Benedetta Flebus}
\email{flebus@bc.edu}
\affiliation{Department of Physics, Boston College, 140 Commonwealth Avenue, Chestnut Hill, Massachusetts 02467, USA}

\begin{abstract}
Several works have recently addressed the emergence of exceptional points (EPs), i.e., spectral singularities of  non-Hermitian Hamiltonians, in the long-wavelength dynamics of coupled magnetic systems. Here, by focusing on the driven magnetization dynamics of a  van der Waals ferromagnetic bilayer,  we show that exceptional points can appear over extended portions of the first Brillouin zone as well. 
Furthermore, we demonstrate that the effective non-Hermitian magnon Hamiltonian, whose eigenvalues are purely real or come in complex-conjugate pairs, respects an unusual wavevector-dependent pseudo-Hermiticity. Finally, for both armchair and zigzag nanoribbon geometries, we discuss  both the complex and purely real spectra of the topological  edge states and their experimental implications.
\end{abstract}

\maketitle

%%%%%%%%%%%%%%%%%%%%%%%%%%%%Main Body%%%%%%%%%%%%%%%%%%%%%%%%%%%%%%%%%%%%%

\section{Introduction}
%After the quantum mechanics based on Hermitian assumption established\cite{schrodinger1926quantisierung, mehra1987erwin-I, mehra1987erwin-II,fedak20091925}, generalizing it to non-Hermitian field had been an important topic from both theoretical and experimental perspectives\cite{moiseyev2011non, bender2007making, ashida2020non}. A Hermitian system has a purely real spectrum and the corresponding non-degenerate eigenvectors are orthogonal\cite{dirac2001lectures,griffiths2018introduction,ballentine2014quantum,merzbacher1961quantum,messiah2014quantum,jammer1974philosophy}, while for  a non-Hermitian one, it possesses a complex spectrum and the non-degenerate eigenvectors satisfy the bi-orthogonal relation\cite{bender2007making,brody2013biorthogonal,curtright2007biorthogonal,kunst2018biorthogonal}, accordingly. 
Non-Hermitian phenomena are ubiquitous in nature and in recent  years have attracted widespread attention in several areas of physics \cite{wiersig2020review,okuma2022non,cao2015dielectric,lee2014heralded,moiseyev2011non,flebus2020non,peng2016chiral,ding2016emergence,liu2019second,el2019dawn,bergholtz2021exceptional,pan2020non,kunst2018biorthogonal,zhao2019non,li2020critical}. While non-Hermitian Hamiltonians  arise in any open system wherein coupling to the environment leads to dissipation, intense research efforts have been devoted to  non-Hermitian phases stemming from the interplay between gain and loss. In particular, whereas loss and gain are balanced, a system might be described by a $\mathcal{P}\mathcal{T}$-symmetric~\cite{brody2016consistency,bender1999pt,bender2005introduction,mostafazadeh2002pseudoI,mostafazadeh2002pseudoII} or pseudo-Hermitian~\cite{mostafazadeh2002pseudoI,mostafazadeh2002pseudoII,bagchi2002pseudo,mostafazadeh2010conceptual,mostafazadeh2006weak,bagchi2005pseudo}  Hamiltonian. The later, despite being non-conservative,  can  display  a purely real energy spectrum due to the existence of stationary states\cite{mostafazadeh2002pseudo}.  In $\mathcal{P}\mathcal{T}$-symmetric and pseudo-Hermitian systems, the eigenvalues undergo a transition from real to complex in correspondence of non-Hermitian spectral degeneracies, i.e., exceptional points (EPs)~\cite{miri2019exceptional,   tserkovnyak2020exceptional,el2018non}.

EPs are spectral singularities in the parameter space
at which both the eigenvalues and eigenvectors coalesce~\cite{heiss2012physics,miri2019exceptional,miri2019exceptional,hodaei2017enhanced,ozdemir2019parity}: their properties are strikingly different from the Hermitian degeneracies at which only eigenvalues degenerate. Generally, $N$ eigenvectors can coalesce at the same EP,
dubbed therefore an $N$-th order EP~\cite{hodaei2017enhanced,mandal2021symmetry,yu2020higher,zhou2018optical}.
For a small deviation $\epsilon\ll1$ around an $N$-th order EP,
 the eigenvalues can be expanded in terms of a Puiseux series
of order $\epsilon^{1/N}\gg\epsilon$. The $N$-th root singularity signals a drastic response of the system to a  perturbation around the EP~\cite{miri2019exceptional,hodaei2017enhanced}, which has led to numerous efforts in the development of EP-based sensors~\cite{wiersig2020review,yu2020experimental,zhong2020exceptional}. 

EPs yield several other intriguing phenomena, such as
 enhanced  spontaneous emission~\cite{lin2016enhanced},  anomalous lasing behavior~\cite{feng2014single,miri2012large,liertzer2012pump,peng2014loss,peng2014loss,takata2021observing}, unidirectional reflectionless light propagation~\cite{huang2017unidirectional,lin2011unidirectional,yin2013unidirectional}, and dynamical phase transitions in nonlinear systems~\cite{deng2022exceptional}.
Heretofore, EPs have been predicted and  observed in a plethora of systems, including but  not limited to photonic~\cite{miri2019exceptional,ozdemir2019parity,zhang2020pt}, magnonic~\cite{liu2019observation,lee2015macroscopic,zhang2019experimental}, electronic ~\cite{dong2019sensitive,stehmann2004observation}, acoustic~\cite{shen2018synthetic,zhu2018simultaneous,ding2016emergence,shi2016accessing}, optical~\cite{peng2016chiral,goldzak2018light,jing2017high,longhi2014optical,ghosh2016exceptional} and magnetic systems~\cite{yu2020higher,liu2019observation,liu2019observation,gunnink2022nonlinear,wang2021enhanced,flebus2020non,wang2018coalescence}.

In magnetic systems,  loss of magnetization dynamics is unavoidable due to ubiquitous interactions between magnetic moments and  crystalline lattice \cite{dai2000magnon,agrawal2013direct,kormann2014temperature,pincus1961influence,deng2022non}, while gain can be introduced via experimentally-established techniques, such as  spin-torque
transfer~\cite{slonczewski1996current,berger1996emission,tsoi1998excitation,myers1999current} or  parametric driving~\cite{tabuchi2015coherent,chen2017parametric}. The feasibility with which the balance between gain and loss can be tuned  makes magnetic systems promising solid-state hosts of   EPs~\cite{yu2020higher,liu2019observation}. 
However, so far the majority of  systems that have been  investigated display second- or higher-order EPs at a single wavevector $\mathbf{k}$ - most commonly at $|\mathbf{k}|=0$, corresponding to the long-wavelength limit of magnetization dynamics. While such EPs can be detected indirectly by probing the system's parameters~\cite{liu2019observation}, it is unlikely that the presence of a single EP will influence the response functions routinely probed in spintronics setups. 
Several system's properties, such as, e.g., transport coefficients, depend on integrals over the entire first Brillouin zone:  a singularity at a single wavevector  will  likely not  significantly affect  their values. 

\begin{figure}[h!]
\includegraphics[width=1\linewidth]{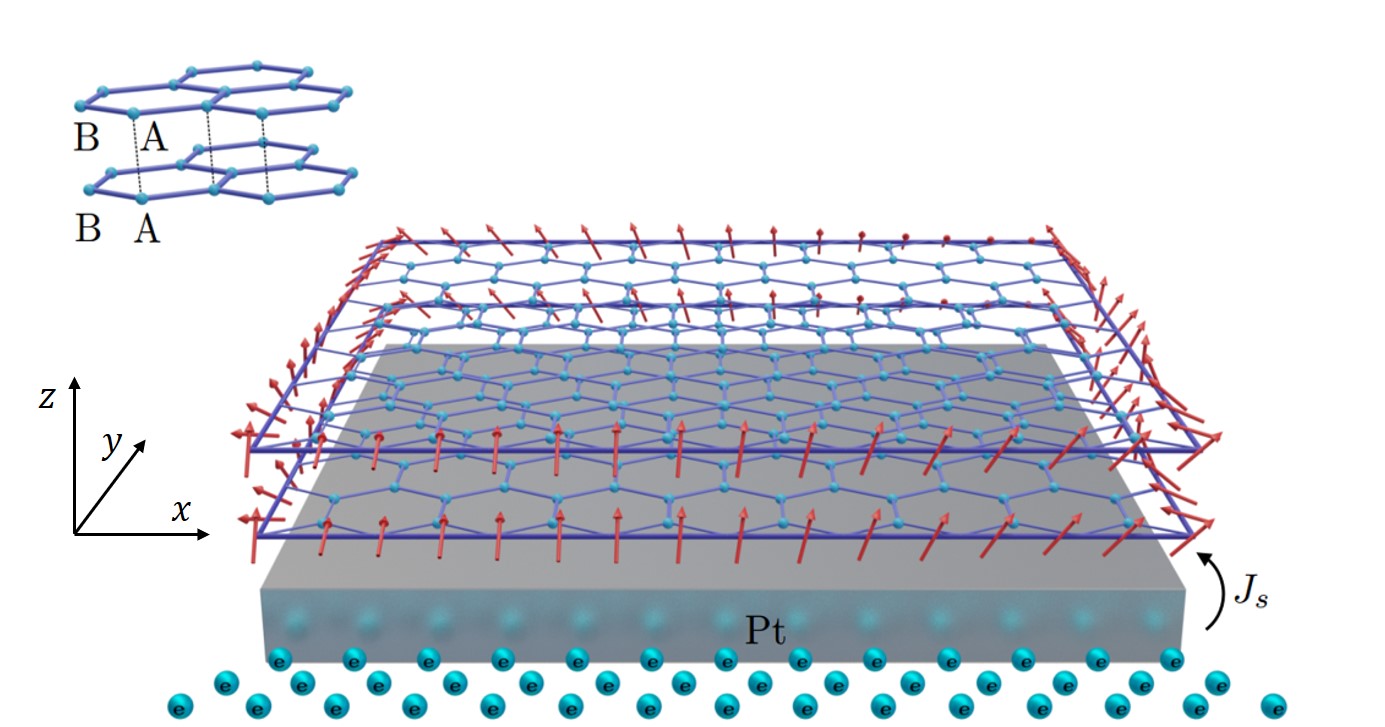}
\caption{Our proposed setup: an AA-stacked vdW ferromagnetic bilayer  deposited on a platinum strip. An electric current flowing into the platinum (Pt) film is converted into a spin current $J_{s}$ injected into the bottom magnetic layer. }\label{material} 
\end{figure} 

In this work, we unveil the emergence of broad patterns of second-order EPs in the reciprocal space of a van der Waals (vdW) bilayer deposited on a platinum substrate, depicted in Fig.~\ref{material}.  Via spin current injection from the platinum substrate  and modulation of the interlayer coupling, one might realize a balance of gain and loss between the bottom and top layer~\cite{tserkovnyak2002enhanced,cheng2014dynamics,shikoh2013spin}.
The resulting spin dynamics can be captured by the coupled
Landau–Lifshitz–Gilbert (LLG) equations~\cite{gilbert1955lagrangian,lakshmanan2011fascinating}, which, upon linearization, yield an effective non-Hermitian Hamiltonian. For balanced gain and loss, we find that the eigenenergies are
always purely real or  come in complex-conjugate pairs. While the spectrum 
is evocative of parity-time ($\mathcal{PT}$-) symmetry or global (i.e., parameter-independent) pseudo-Hermiticity, we show  that  our
model possesses instead a $\mathbf{k}$-dependent pseudo-Hermiticity. We find that a multitude of exceptional points - arranged in, e.g.,  exceptional rings~\cite{xu2017weyl,zhen2015spawning} - can appear in the first Brillouin zone of the honeycomb lattice.
Finally, we investigate the topological edge states that emerge in the presence of  Dyzaloshinskii–Moriya interaction (DMI) and reveal topological insulating phases analogous to Hermitian quantum theories.

 This work is organized as follows: In Sec. II we derive the effective non-Hermitian Hamiltonian describing our model. In Sec. III,  we perfom a symmetry analysis of the non-Hermitian Hamiltonian.  In Sec. IV, we discuss the emergence of multitude of exceptional points and topological edge states.  In Sec. V we present our conclusions and an outlook.

\section{Model}

 We consider an AA-stacked  ferromagnetic bilayer with spins localized on the honeycomb lattice. The Hamiltonian is given as

\begin{eqnarray}
\mathcal{H}_0&=&-J\sum_{\langle i,j\rangle}\mathbf{m}^{(l)}_i\cdot\mathbf{m}^{(l)}_j-\lambda\sum_i\mathbf{m}_i^{(1)}\cdot\mathbf{m}_i^{(2)}\nonumber\\
&&-A\sum_{i}(\mathbf{m}_i^{(l)}\cdot \hat{\mathbf{z}})^2-M_s\sum_i\mathbf{m}^{(l)}_i\cdot\mathbf{h}\nonumber\\
&&+D\sum_{\langle\langle i,j\rangle\rangle}\nu_{ij}\hat{\mathbf{z}}\cdot(\mathbf{m}_i^{(l)}\times\mathbf{m}_j^{(l)}),
\label{96}
\end{eqnarray}
where $\mathbf{m}^{(l)}_i$ is the  magnetic moment orientation at the $i$th site of the $l$th layer, and $l=1,2$ labels the bottom and top layers, respectively. $J>0$  parameterizes the ferromagnetic nearest-neighbor (NN) Heisenberg interaction, $D>0$ the next-to-nearest neighbors (NNN) Dzyaloshinskii-Moriya interaction  (DMI), with $\nu_{ij}=-\nu_{ji}=\pm1$, $A>0$ the uniaxial anisotropy,  $\lambda>0$ the ferromagnetic interlayer coupling exchange strength, $M_s>0$ the magnitude of magnetic moments, and $\mathbf{h}=h\hat{\mathbf{z}}$ an external magnetic field oriented along the $z$ direction.
In our proposed setup, shown in Fig.~\ref{material}, a charge current flowing into a metal with strong spin-orbit coupling results into  injection of a spin current $J_{s}$ into the bottom layer via the spin Hall effect \cite{hirsch1999spin,sinova2004universal,valenzuela2006direct}. We assume the current to be strong enough to compensate for the damping of magnetization dynamics in the bottom layer, yielding effective gain~\cite{flebus2020non}. 
The ferromagnetic layers interact via weak van der Waals forces, whose strength can be further tuned by, e.g., adding a non-magnetic layer within them. Here we focus on a  state with balanced gain and loss, i.e., we assume that the interlayer coupling is weak enough that only a small portion of the current $J_{s}$ reaches the top layer, whose dynamics remains then lossy.

The dynamics of the magnetization $\mathbf{m}^{(1,2)}_{j}$ at the $j$th site of the bottom (top) layer obey the LLG equations:

\begin{eqnarray}
\dot{\mathbf{m}}^{(1)}_{j}&=&-\gamma\mathbf{m}^{(1)}_j\times \mathbf{h}^{(1)}_{j,\text{eff}}-\alpha\mathbf{m}^{(1)}_j\times\dot{\mathbf{m}}^{(1)}_{j},
\nonumber\\
\dot{\mathbf{m}}^{(2)}_{j}&=&-\gamma\mathbf{m}^{(2)}_j\times \mathbf{h}^{(2)}_{j,\text{eff}}+\alpha\mathbf{m}^{(2)}_j\times\dot{\mathbf{m}}^{(2)}_{j}, \label{2}
\end{eqnarray}
where $\dot{\mathbf{m}}^{(l)}_j= \frac{d \mathbf{m}^{(l)}_j}{ dt}$.
Here $\gamma>0$ is the  gyromagnetic ratio, $\alpha>0$ the Gilbert damping,  and we have introduced the effective magnetic field $\mathbf{h}^{(l)}_{j,\text{eff}}=-M_s^{-1}\partial \mathcal{H}_0/\partial \mathbf{m}^{(l)}_j$.
Substituting  Eq.~\eqref{2} again for $\dot{\mathbf{m}}^{(l)}_{j}$ in the third term of Eq.~\eqref{2} yields

\begin{eqnarray}
\dot{\mathbf{m}}^{(1)}_{j}
&=&\frac{-\gamma}{1+\alpha^2}\Big[\mathbf{m}_j^{(1)}\times \mathbf{h}^{(1)}_{j,\text{eff}}
-\alpha\mathbf{m}_j^{(1)}\times \left(\mathbf{m}_j^{(1)}\times \mathbf{h}_{j,\text{eff}}^{(1)}\right)\Big],\nonumber\\
 \dot{\mathbf{m}}^{(2)}_{j}
&=&\frac{-\gamma}{1+\alpha^2}\Big[\mathbf{m}_j^{(2)}\times \mathbf{h}^{(2)}_{j,\text{eff}}
+\alpha\mathbf{m}_j^{(2)}\times \left( \mathbf{m}_j^{(2)}\times \mathbf{h}_{j,\text{eff}}^{(2)}\right)\Big].\nonumber\\
\label{LLGoriginal}
\end{eqnarray}
To obtain the effective non-Hermitian Hamiltonian associated with Eq.~(\ref{LLGoriginal}), we consider small deviations of the magnetic moment from its equilibrium direction, here taken to be along $\hat{\mathbf{z}}$, by assuming $\mathbf{m}_{i}^{(l)}\simeq(\delta m_i^{(l),x},\delta m_i^{(l),y},1)$, with
\begin{eqnarray}
\delta m_i^{(l),x(y)}&=&|\delta m_i^{(l),x(y)}|e^{i (\omega t-\mathbf{k}\cdot\mathbf{r}_{i})},
\label{131}
\end{eqnarray}
where $\omega$ and $\mathbf{k}=(k_x,k_y)$ are the spin-wave frequency and wavevector, respectively, and $\mathbf{r}_{i}$  denotes the position of the $i$th magnetic moment. By plugging Eq.~(\ref{131}) into the linearized form of Eq.~(\ref{LLGoriginal}),  we find

\begin{eqnarray}
\mathcal{H}(\mathbf{k})\Psi(\mathbf{k})=E(\mathbf{k})\Psi(\mathbf{k}),
\end{eqnarray}
where   $\Psi(\mathbf{k})=(\psi^{(1)}_A(\mathbf{k}),\psi^{(1)}_B(\mathbf{k}),\psi^{(2)}_A(\mathbf{k}),\psi^{(2)}_B(\mathbf{k}))^T$, with

\begin{eqnarray}
\psi^{(l)}_A(\mathbf{k})&=&\delta m_{A,\mathbf{k}}^{(l),x}+i\delta m_{A,\mathbf{k}}^{(l),y},\nonumber\\
\psi^{(l)}_B(\mathbf{k})&=&\delta m_{B,\mathbf{k}}^{(l),x}+i\delta m_{B,\mathbf{k}}^{(l),y}.
\end{eqnarray}
and $E(\mathbf{k})=\hbar\omega(\mathbf{k})$. The effective non-Hermitian Hamiltonian $\mathcal{H}(\mathbf{k})$ reads  as

\begin{widetext}
\begin{eqnarray}
\mathcal{H}(\mathbf{k})
&=&\frac{\gamma\hbar}{M_s(1+\alpha^2)}\left(
\begin{array}{cccc}
(1+i\alpha)(C-2D\Delta_{\mathbf{k}})&-(1+i\alpha)J\gamma_{\mathbf{k}}&-(1+i\alpha)\lambda&0\\
-(1+i\alpha)J\gamma^*_{\mathbf{k}}&(1+i\alpha)(C+2D\Delta_{\mathbf{k}})&0&-(1+i\alpha)\lambda\\
-(1-i\alpha )\lambda&0&(1-i\alpha )(C-2D\Delta_{\mathbf{k}})&-(1-i\alpha )J\gamma_{\mathbf{k}}\\
0&-(1-i\alpha )\lambda&-(1-i\alpha )J\gamma^*_{\mathbf{k}}&(1-i\alpha )(C+2D\Delta_{\mathbf{k}})
\end{array} 
\right)\label{effnonhermitian}
\end{eqnarray}
\end{widetext}
The parameters introduced in Eq.~(\ref{effnonhermitian}) are given as 
\begin{align}
C&=3J+2A+\lambda+hM_s,\\
\gamma_{\mathbf{k}}&=e^{-i\mathbf{k}\cdot\delta_1}+e^{-i\mathbf{k}\cdot\delta_2}+e^{-i\mathbf{k}\cdot\delta_3}, \\
\gamma^*_{\mathbf{k}}&=e^{i\mathbf{k}\cdot\delta_1}+e^{i\mathbf{k}\cdot\delta_2}+e^{i\mathbf{k}\cdot\delta_3},\\
\Delta_{\mathbf{k}}
&=-\sin\mathbf{k}\cdot\gamma_1+\sin\mathbf{k}\cdot\gamma_2-\sin\mathbf{k}\cdot\gamma_3,
\end{align}
where the NN and NNN relative position vectors $\delta_i$ and $\gamma_i$ (i=1,2,3) are

\begin{eqnarray}
\delta_1=\Big(\frac{\sqrt{3}}{2},-\frac{1}{2}\Big)a,
&&\qquad \delta_2=\Big(-\frac{\sqrt{3}}{2},-\frac{1}{2}\Big)a, \nonumber \\
\delta_3=(0,1)a,
&&\qquad\gamma_1=(\sqrt{3},0)a,\nonumber \\
\gamma_2=\Big(\frac{\sqrt{3}}{2},\frac{3}{2}\Big)a,
&&\qquad\gamma_3=\Big(-\frac{\sqrt{3}}{2},\frac{3}{2}\Big)a,
\end{eqnarray}
with $a$  the honeycomb lattice constant.

The Hamiltonian~(\ref{effnonhermitian})  is diagonalized by  a complete set of biorthonormal eigenvectors $\{ |\phi^R\rangle, |\phi^L\rangle\}$, i.e.,
\begin{align}
\mathcal{H}|\phi^R_n\rangle&=E_n|\phi^R_n\rangle,  \; \; \; \; \; \;
\mathcal{H}^\dag|\phi^L_n\rangle=E^*_n|\phi^L_n\rangle,
\end{align}
where  $E_{n}$ is the eigenenergy of the $n$th band, while the right, $ |\phi^R\rangle$,  and left, $|\phi^L\rangle$,  eigenvectors   satisfy

\begin{eqnarray}
\langle\phi^L_n|\phi_m^R\rangle=\delta_{mn},
\qquad\sum_n|\phi_n^R\rangle\langle\phi^L_n|=1\,,
\end{eqnarray} 
with $n,m=1, ..,4$.

\begin{figure}[t!]
   \includegraphics[width=\linewidth]{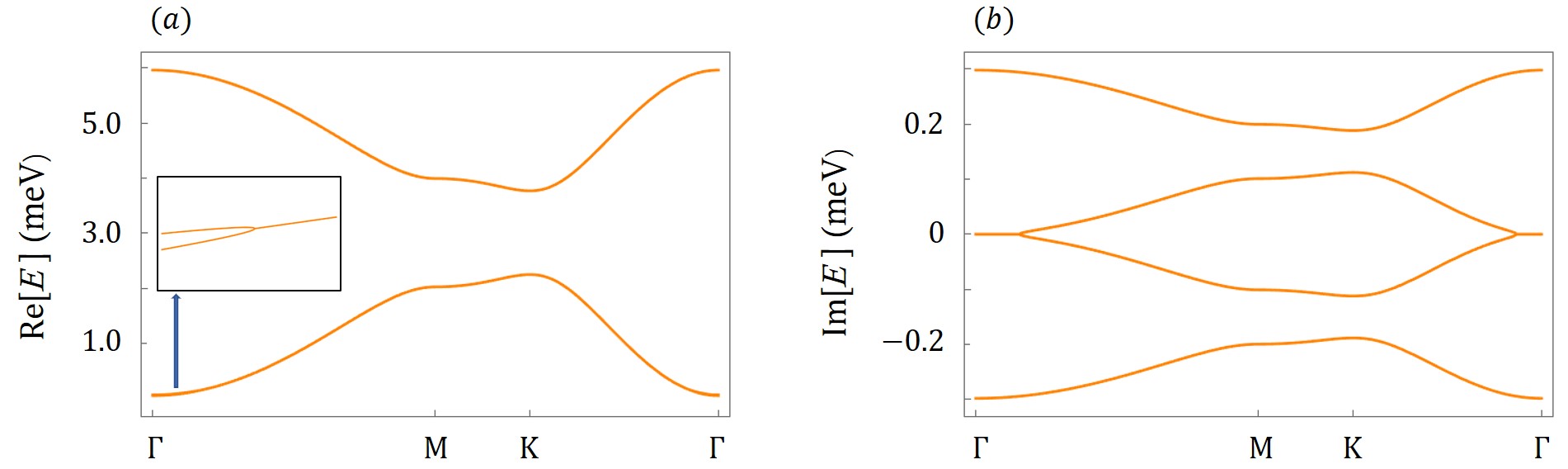}
     \includegraphics[width=\linewidth]{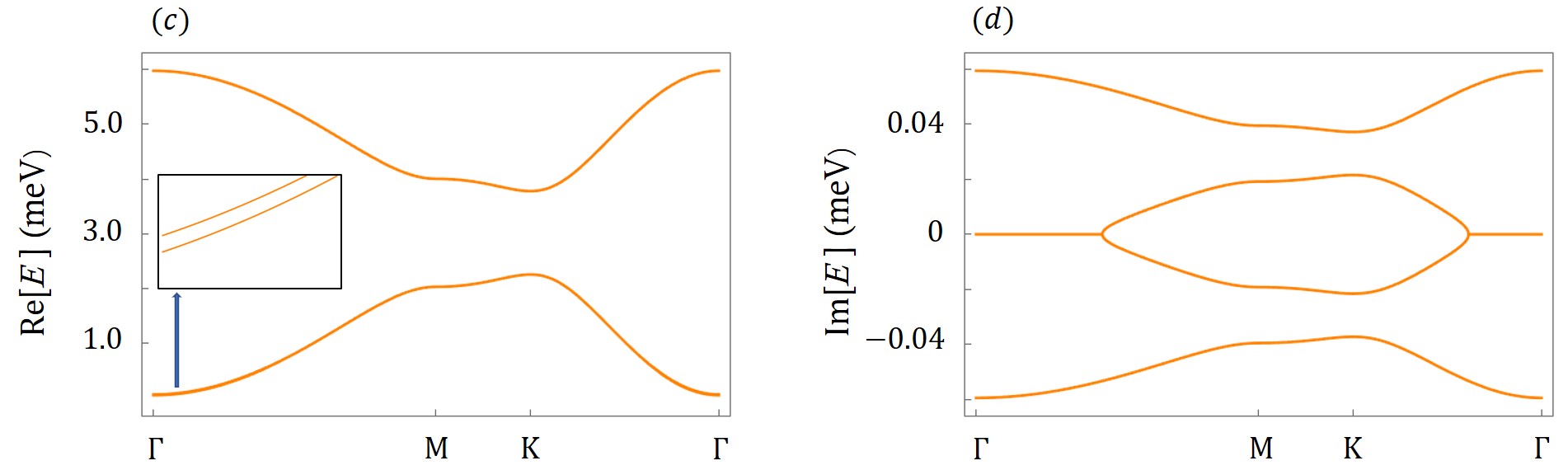}
    \includegraphics[width=\linewidth]{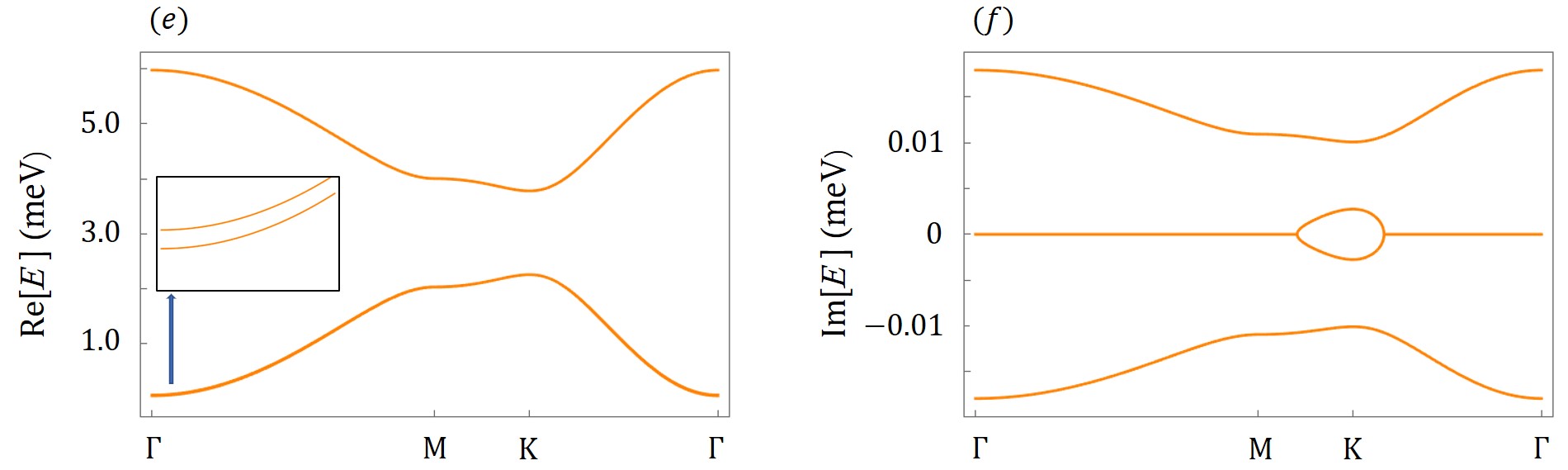}
    \includegraphics[width=\linewidth]{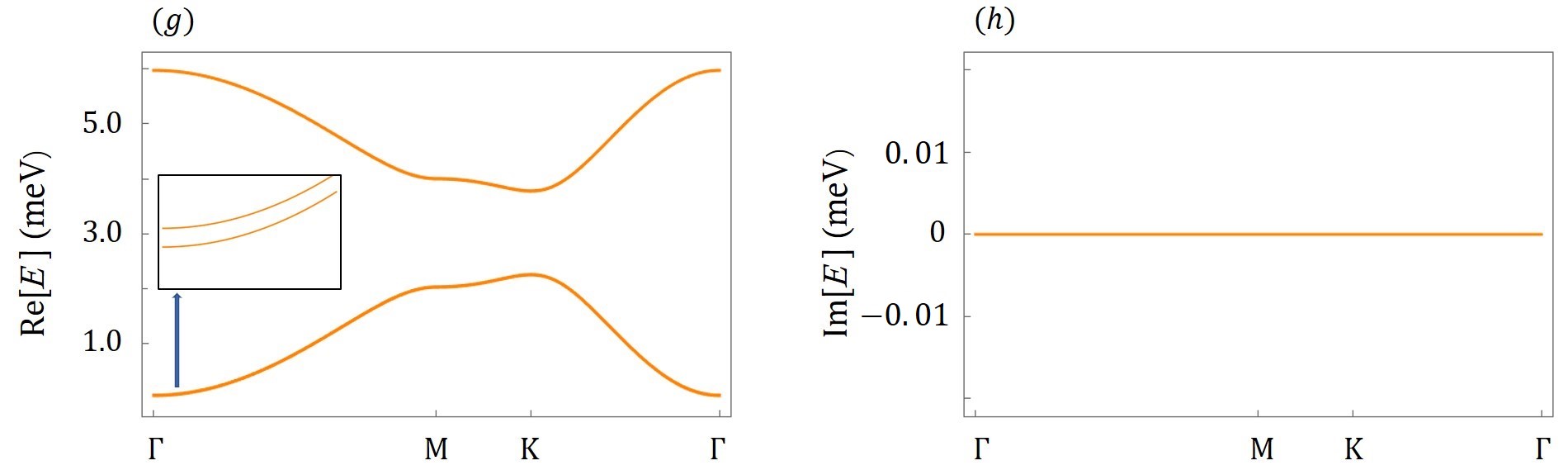}
    \caption{The real (left panels) and imaginary (right panels) magnonic energy spectra along  high-symmetry directions for different values of the gain/loss parameter $\alpha$. From the top to the bottom,  the degree of non-Hermiticity is varied as $\alpha=0.05 \; \text{(a,b)},\  0.01 \; \text{(c,d)},\  0.0032 \; \text{(e,f)},\  0.001 \; \text{(g,h)}$. Other parameters are set to $J=1.48$ meV, $ A=0.02$ meV, $D=0.22$ meV,  $\lambda=0.01$ meV, $M_{s}=3\mu_{B}$ and $h=0.1$ T.}
    \label{Spectra:highsymmetry}
  \end{figure}

Figures~\ref{Spectra:highsymmetry} show the complex energy spectra of the Hamiltonian~(\ref{effnonhermitian}) for different values of $\alpha$.  With a CrBr$_3$ bilayer in mind, we take $  J=1.48$ meV, $A=0.02$ meV, $D=0.22$ meV, $\lambda=0.01$ meV~\cite{cai2021topological}, $M_s=3\mu_B$, with $\mu_B$  the Bohr magneton, and set $h=0.1$ T. We consider damping coefficients of the same order of magnitude of the ones reported for chromium trihalide crystals, i.e., $\alpha \sim 0.01-0.001$~\cite{shen2021multi}. 
The real energy spectrum displays four bands: the two lower (higher)-energy bands correspond to the acoustic (optical)  modes of the top and bottom layers, whose degeneracy is lifted by the interlayer coupling $\lambda$, as shown in  Figs.~\ref{Spectra:highsymmetry}(a),(c),(e) and~(g).  The  gap between the two lower- and the two-higher energy bands is controlled by the strength of the DMI.
From Figs.~\ref{Spectra:highsymmetry}(b),(d),(f) and~(h), one can easily see that the eigenvalues are purely real or come in complex-conjugate pairs depending on the strength of the parameter $\alpha$ introducing non-Hermiticity. Besides, the real and the corresponding imaginary levels always avoid crossing simultaneously, except at EPs~\cite{heiss2000repulsion}.

Eigenvalues being either real or coming in complex-conjugate pairs suggests that the Hamiltonian~\eqref{effnonhermitian} is $\mathcal{PT}$-symmetric or pseudo-Hermitian. As we will discuss in more details in the next section, our Hamiltonian holds a peculiar type of pseudo-Hermiticity, which is $\mathbf{k}$-dependent.

\section{Symmetry analysis\label{Sec:symmetry}}

Hamiltonians exhibiting eigenvalues that are purely real or come into complex-conjugate pairs usually fall into two categories, i.e., $\mathcal{P}\mathcal{T}$-symmetric or pseudo-Hermitian. A Hamiltonian ~$\mathcal{H}$ is dubbed $\mathcal{PT}$-symmetric if it satisfies the relation~\cite{kawabata2019symmetry}
\begin{eqnarray}
\mathcal{M}\mathcal{H}^*(\mathbf{k})\mathcal{M}^{-1}=\mathcal{H}(\mathbf{k}),
\label{PTcondition}
\end{eqnarray} 
in reciprocal space, where $\mathcal{M}$ is a unitary matrix, with $\mathcal{M}\mathcal{M}^*=\pm1$. On the other hand, pseudo-Hermiticity  requires~\cite{kawabata2019symmetry}

\begin{eqnarray}
\eta \mathcal{H}^{\dag}(\mathbf{k})\eta^{-1}=\mathcal{H}(\mathbf{k}),
\label{PseudoHermicitycondition}
\end{eqnarray}
where  $\eta$ is a unitary and Hermitian matrix, with $\eta^2=1$.  When $\eta=1$, Eq.~(\ref{PseudoHermicitycondition})  reduces to the condition of Hermiticity. 

We find that there is no parameter-independent linear automorphism such that the  Hamiltonian~(\ref{effnonhermitian}) satisfies (\ref{PseudoHermicitycondition}) for every wavevector $\mathbf{k}$. To prove it, it is convenient to rewrite the Hamiltonian~\eqref{effnonhermitian} - normalized  by $\hbar \gamma / M_s(1+\alpha^2)$ - as
\begin{eqnarray}
\mathcal{H}(\mathbf{k})&=&
-\frac{1}{2}J\gamma_{\mathbf{k}}(\sigma_0\otimes\sigma_1-\alpha \sigma_3\otimes\sigma_2)\nonumber\\
&&-\frac{1}{2}J\gamma^*_{\mathbf{k}}(\sigma_0\otimes\sigma_1+\alpha \sigma_3\otimes\sigma_2)\nonumber\\
&&-\frac{1}{2}i J\gamma_{\mathbf{k}}(\sigma_0\otimes\sigma_2+\alpha\sigma_3\otimes\sigma_1)\nonumber\\
&&+\frac{1}{2}i J\gamma^*_{\mathbf{k}}(\sigma_0\otimes\sigma_2-\alpha\sigma_3\otimes\sigma_1)\nonumber\\
&&-2D\Delta_{\mathbf{k}}(\sigma_0\otimes\sigma_3+i\alpha\sigma_3\otimes\sigma_3)\nonumber\\
&&-\lambda(\sigma_1\otimes\sigma_0-\alpha\sigma_2\otimes\sigma_0)\nonumber\\
&&+C(\sigma_0\otimes\sigma_0+i\alpha\sigma_3\otimes\sigma_0),
\nonumber\\ \label{227}
\end{eqnarray}  
where $\sigma_{0}$ represents the $2\times2$ identity matrix, and $\sigma_{i}$ is the $i$th Pauli matrix with $i=1, 2, 3$. Let us assume that the Hamiltonian has global pseudo-Hermiticity with

\begin{eqnarray}
\eta=\sum_{\mu,\nu} a_{\mu\nu}\sigma_\mu\otimes \sigma_\nu,\quad \; \; \;  \mu,\nu=0,1,2,3,
\end{eqnarray}
where $a_{\mu\nu}$ are constants. Equation~(\ref{PseudoHermicitycondition}) yields
\begin{eqnarray}
\sum_{\mu,\nu} a_{\mu\nu}\sigma_\mu\otimes \sigma_\nu \mathcal{H}(\mathbf{k})=\sum_{\mu,\nu} a_{\mu\nu}\mathcal{H}^\dag(\mathbf{k})\sigma_{\mu}\otimes \sigma_{\nu}. 
\label{238}
\end{eqnarray}
Taking the term  $\propto D$ in Eq.~(\ref{227}) as an example, one can show that the constraint~(\ref{238}) leads to
\begin{align}
&\sum_{\mu,\nu} a_{\mu\nu}\sigma_\mu\otimes \sigma_\nu (\sigma_0\otimes\sigma_3+i\alpha\sigma_3\otimes\sigma_3)\nonumber\\
&=\sum_{\mu,\nu} a_{\mu\nu}(\sigma_0\otimes\sigma_3-i\alpha\sigma_3\otimes\sigma_3)\sigma_{\mu}\otimes \sigma_{\nu},
\end{align}
which implies
\begin{align}
a_{00}=a_{01}=a_{02}=a_{03}=a_{11}=a_{22}=0,\\
a_{12}=a_{21}=a_{30}=a_{33}=
a_{31}=a_{32}=0.
\end{align}
Similarly, by considering the other terms entering Eq.~(\ref{227}), we find that Eq.~(\ref{238}) is satisfied only when

\begin{eqnarray}
a_{\mu\nu}=0 \quad\text{for}\quad \mu,\nu=0,1,2,3,
\end{eqnarray}
indicating that the Hamiltonian can not have a  $\mathbf{k}$-independent (i.e., global) pseudo-Hermiticity~\eqref{PseudoHermicitycondition} for all values of the momenta. Following the same approach, it is straightforward to prove that the Hamiltonian~\eqref{effnonhermitian}  does not possess $\mathcal{P}\mathcal{T}$-symmetry~\eqref{PTcondition} either. 

We find that $\mathcal{PT}$-symmetry and global pseudo-Hermiticity are restored only in the long-wavelength limit, i.e., by setting $\mathbf{k}=0$ in Eq.~(\ref{effnonhermitian}).
For vanishing DMI strength, i.e., $D=0$, one recovers $\mathcal{PT}$-symmetry over the full Brillouin zone,  but no global pseudo-Hermiticity. We have also explored an antiferromagnetically coupled ferromagnet bilayer, realized by our model~(\ref{96}) for $\lambda<0$, and found that it can not display $\mathcal{PT}$-symmetry  or global pseudo-Hermiticity, independently of the values of the wave vector $\mathbf{k}$ and DMI strength $D$.

 To explain the purely real or complex conjugate paired spectrum of Eq.~(\ref{effnonhermitian}), one needs instead to invoke a generalized pseudo-Hermiticity 
\begin{align}
\eta(\xi) \mathcal{H}(\xi)^{\dag}\eta^{-1}(\xi)=\mathcal{H}(\xi)\ ,
\label{273}
\end{align}
where the unitary Hermitian matrix $\eta$ depends continuously on a parameter $\xi$. In our case, when setting $\xi \rightarrow \mathbf{k}$, Eq.~(\ref{273}) holds for every wavevector $\mathbf{k}$.
The matrix $\eta(\mathbf{k})$ can be constructed from eigenbasis of $\mathcal{H}(\mathbf{k})$~\cite{mostafazadeh2002pseudo}, i.e.,

\begin{eqnarray}
\eta(\mathbf{k})&=&+|\phi^R_1(\mathbf{k})\rangle\langle \phi^R_2(\mathbf{k})|+|\phi^R_2(\mathbf{k})\rangle\langle \phi^R_1(\mathbf{k})|\nonumber\\
&&+|\phi^R_3(\mathbf{k})\rangle\langle \phi^R_4(\mathbf{k})|+|\phi^R_4(\mathbf{k})\rangle\langle \phi^R_3(\mathbf{k})|,
\end{eqnarray}

or

\begin{eqnarray}
\eta^{-1}(\mathbf{k})&=&+|\phi^L_1(\mathbf{k})\rangle\langle \phi^L_2(\mathbf{k})|+|\phi^L_2(\mathbf{k})\rangle\langle \phi^L_1(\mathbf{k})|\nonumber\\
&&+|\phi^L_3(\mathbf{k})\rangle\langle \phi^L_4(\mathbf{k})|+|\phi^L_4(\mathbf{k})\rangle\langle \phi^L_3(\mathbf{k})|.
\end{eqnarray}

\begin{figure*}[htbp]
    \centering
    \includegraphics[width=0.9\linewidth]{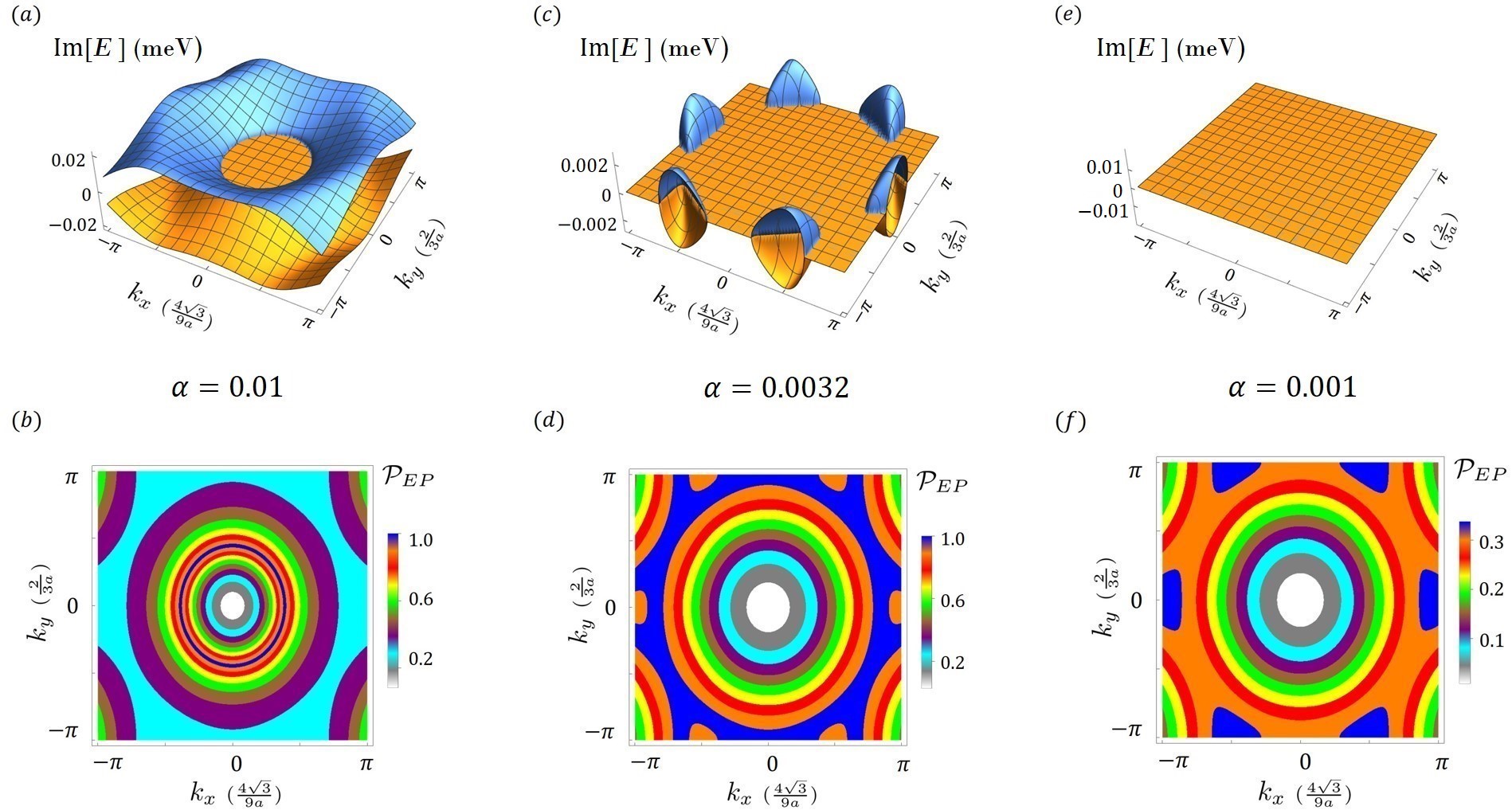}
    \caption{Top panels: imaginary parts of the first, $E_{1}$, and second, $E_{2}$, lowest-energy eigenvalues. Bottom panels:  square of the overlap of the two right eigenvectors, $\mathcal{P}_{EP}\equiv \abs{\bra{\phi_1^R}\ket{\phi_2^R}}^2$, within the first Brillouin zone. The non-Hermiticity is tuned to $\alpha=0.01 \; \text{(a,b)},\ 0.0032 \; (\text{c,d}),\  0.001 \; (\text{e,f})$.
Other parameters are set to $J=1.48$ meV, $ A=0.02$ meV, $D=0.22$ meV,  $\lambda=0.01$ meV, $M_{s}=3\mu_{B}$ and $h=0.1$ T.}
    \label{Fig2}
\end{figure*}

\section{Results}

\subsection{Exceptional points}

In  pseudo-Hermitian systems,  EPs 
signal a symmetry-breaking transition at which a system's 
eigenvalues turn from real to complex-conjugate pairs. It is thus not 
surprising that EPs  emerge in Eq.~\eqref{effnonhermitian} at the 
value of $\alpha$ that separates a complex conjugated spectrum from 
a purely real one. What is remarkable  is the simultaneous emergence of multiple EPs  in the first 
Brillouin zone. Figures~\ref{Fig2}(a), (c) and (e) show the Riemann surfaces of the imaginary parts of the two lowest (real-energy) bands of Eq.~(\ref{effnonhermitian})  for, respectively,   $\alpha=0.01,\  0.0032,\  0.001$.  
For $\alpha=0.01, 0.0032$, both real and complex-conjugate  eigenvalues are present in reciprocal space,  while a weaker non-Hermiticity, i.e., $\alpha=0.001$, allows for a purely real spectrum at every wavevector. Figure~\ref{Fig2}(a) shows that a ring of exceptional points (ER) appears, and, for smaller $\alpha$,  evolves into a pattern with   the hexagonal symmetry of the honeycomb lattice, as shown in Fig.~\ref{Fig2}(c).

It is not straightforward to extrapolate the precise shape of the continuum formed by EPs from the energy spectra displayed in Figs.~\ref{Fig2}(a) and (c). To visualize in more detail the EPs distribution over the first Brillouin zone,  we plot (the square of) the overlap, $\mathcal{P}_{EP}\equiv \abs{\bra{\phi_1^R}\ket{\phi_2^R}}^2$, of  the two lowest-energy right eigenvectors of the non-Hermitian Hamiltonian (\ref{effnonhermitian}) in Figs.~\ref{Fig2}(b), (d) and (f). While approaching an exceptional point, the two eigenstates coalesce, i.e., $\mathcal{P}_{EP} \rightarrow 1$. 

It is natural to expect that EPs dwelling in significant portions of the first Brillouin zone might influence any physical properties whose strength depends on an integration over the all states. 

While we have focused on degeneracies between the two lowest (real-energy) bands, an analysis of the EPs emerging between the two higher-energy bands reveal analogous features.

\subsection{Topological edge states}

The Hermitian Hamiltonian~\eqref{96} is known to be topologically nontrivial~\cite{owerre2016magnon}. The source of a nontrivial band gap is the Dzyaloshinskii-Moriya interaction, which breaks the time-reversal symmetry of the magnon Hamiltonian~\cite{kim2016realization}.
To explore the topology of its non-Hermitian counterpart~(\ref{effnonhermitian}), we  start by introducing a gauge-invariant Berry curvature~\cite{berry1984quantal,vanderbilt2018berry}

\begin{eqnarray}
B^{\mu\nu}_{n,ij}(\mathbf{k})\equiv i\langle\partial_i\phi_n^\mu(\mathbf{k})|\partial_j\phi_n^\nu(\mathbf{k})\rangle,
\label{336}
\end{eqnarray}
where $i,j=k_x, k_y$ and  $\mu, \nu=L,R$ . Invoking Eq.~(\ref{336}), four  Chern numbers can be defined  for each $n$th band as

\begin{eqnarray}
C_n^{\mu\nu}=\frac{1}{2\pi}\int_{BZ}d\mathbf{k}\; [B^{\mu\nu}_{n,k_xk_y}(\mathbf{k})-B^{\mu\nu}_{n,k_yk_x}(\mathbf{k})].
\label{288}
\end{eqnarray}

H. Shen \textit{et al.}~\cite{shen2018topological} have shown that a non-Hermitian Chern topological phase is characterized by a single Chern number, i.e., $C_n^{LL}=C_n^{LR}=C_n^{RL}=C_n^{RR}$. We  calculate $C_n^{RR}$ using Fukui's method~\cite{fukui2005chern} while setting $\alpha=0.01$ and  find  $C_1^{RR}=C_2^{RR}=-C_3^{RR}=-C_4^{RR}=1$.

\begin{figure}
     \includegraphics[width=0.99\linewidth]{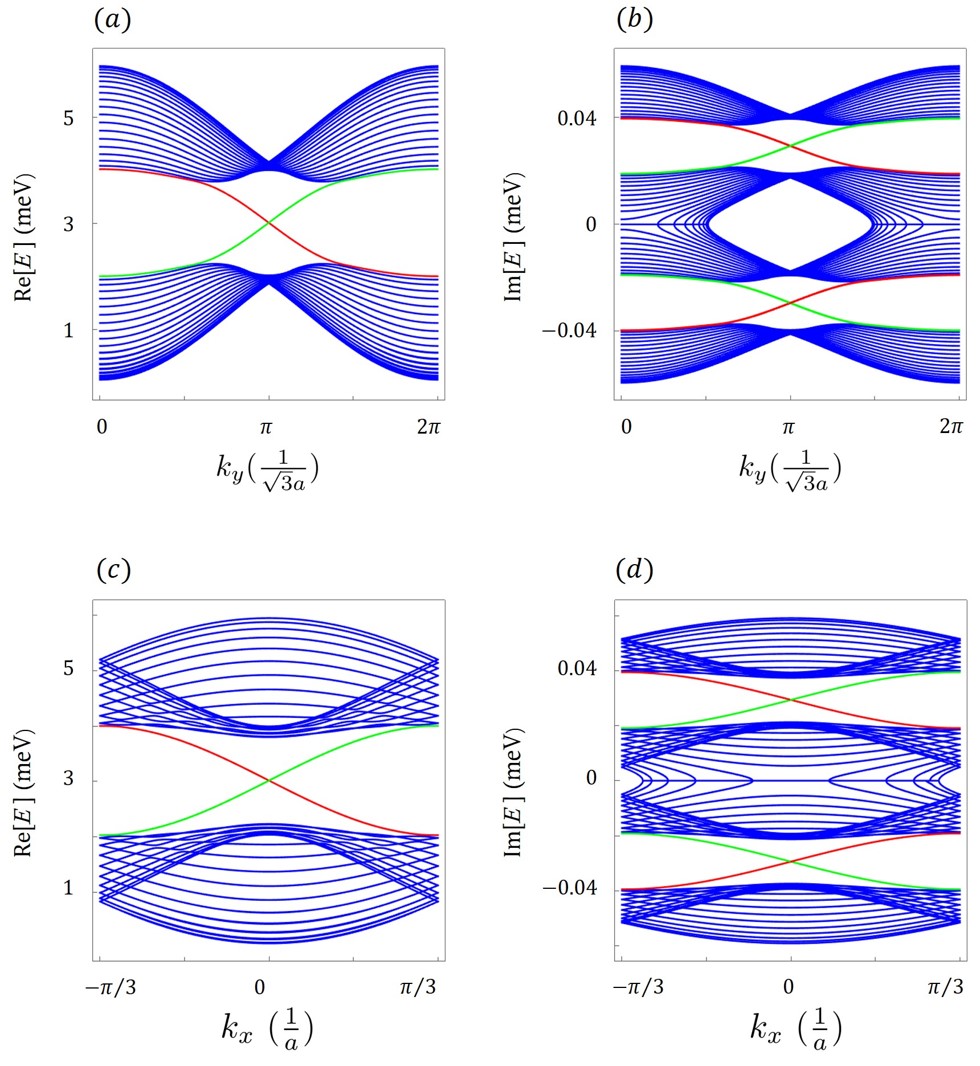}
  \caption{(a,b): Real and imaginary part of the energy spectrum on a zigzag ribbon with open boundary conditions along the $x$ direction. (c,d): Real and imaginary part of the energy spectrum on an armchair ribbon with open boundary conditions along the $y$ direction. Edge states (red and green lines) come in complex-conjugate pairs. Blue lines represent the bulk modes.  The parameters are set to $J=1.48$ meV, $ A=0.02$ meV, $D=0.22$ meV,  $\lambda=0.01$ meV, $M_{s}=3\mu_{B}$, $h=0.1$ T and $\alpha=0.01$.  }\label{OBCspectrum}
\end{figure}

The complex frequency spectra that result from exact diagonalization of Eq.~(\ref{effnonhermitian}) for $\alpha=0.01$ in a zigzag  and armchair open boundary geometry are displayed  in Figs.~\ref{OBCspectrum}(a) and (b), and Figs.~\ref{OBCspectrum}(c) and (d), respectively.  In both geometries, we find four edge states - in agreement with the  Chern numbers~\eqref{288}.  
 Figures~\ref{OBCspectrum}(a) and~(c) show that, in both geometries, the edge states reside in the (real-energy) bulk gap and display a double degeneracy.  On the other hand, due to level repulsion~\cite{heiss2000repulsion}, there is no degeneracy in the imaginary part of the complex energy spectrum. Instead, edge states come in complex-conjugate pairs, as shown in Fig.~\ref{OBCspectrum}(b) and (d).

\begin{figure}
     \includegraphics[width=0.99\linewidth]{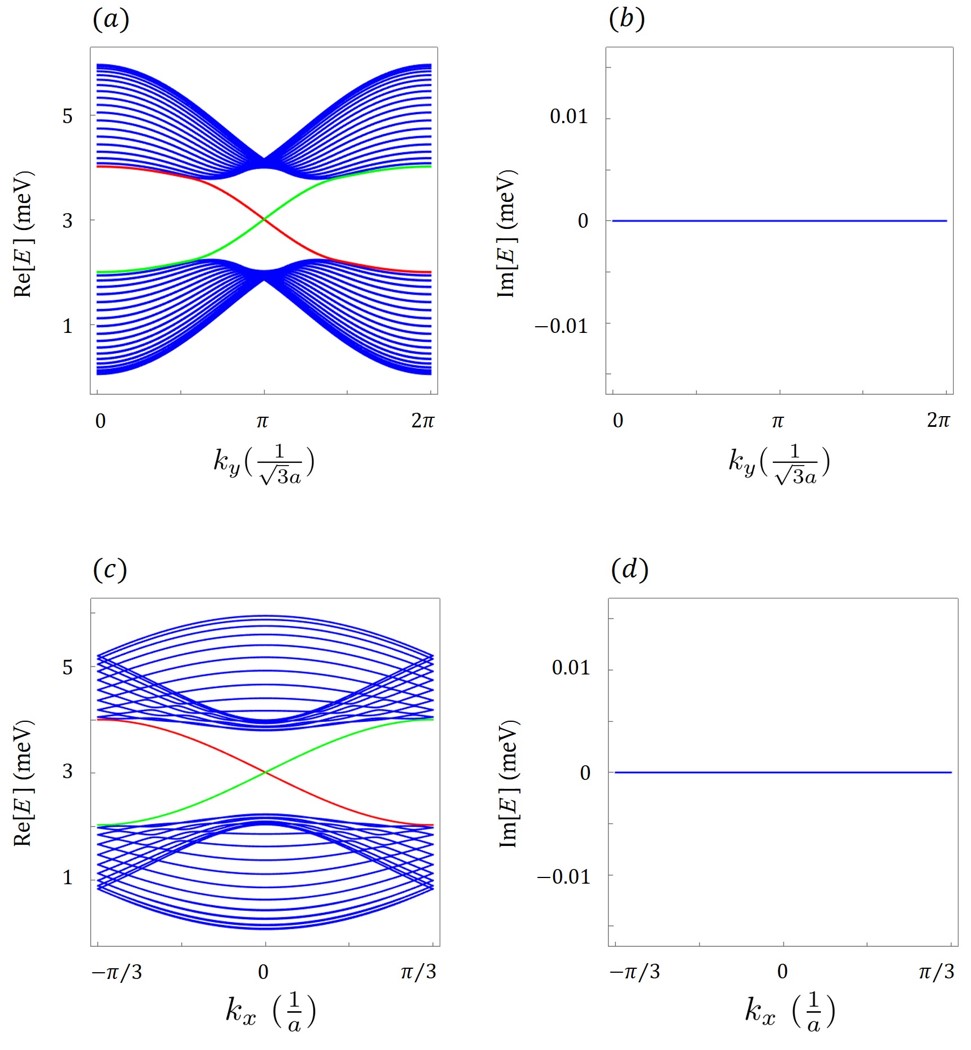}
  \caption{(a,b): Real and imaginary part of the energy spectrum on a zigzag ribbon with open boundary conditions along the $x$ direction. (c,d): Real and imaginary part of the energy spectrum on an armchair ribbon with open boundary conditions along the $y$ direction. Red and green lines correspond to the edge states, while blue lines represent the bulk modes. We find that the spectra are purely real.  The parameters are set to $J=1.48$ meV, $ A=0.02$ meV, $D=0.22$ meV,  $\lambda=0.01$ meV, $M_{s}=3\mu_{B}$, $h=0.1$ T and $\alpha=0.001$.  }\label{OBCspectrumreal}
\end{figure}

Figures~\ref{OBCspectrumreal}(a-d) show that the energy spectra for both zigzag and armchair nanoribbons become purely real when the strength of non-Hermiticity is decreased, i.e., $\alpha=0.001$.
We find that, due to the pseudo-Hermiticity of our model~(\ref{effnonhermitian}), the spectra are always real or complex conjugate paired. Thus, our results suggest that  wave-vector dependent pseudo-Hermiticity~(\ref{273}), contrary to $\mathcal{PT}$-symmetry ~\cite{zhou2018dynamical},  can yield truly topological insulating states. 

However, in a magnetic system,  a purely real bulk and edge spectrum might hinder the direct observation of  topological edge state due to the Bose-Einstein occupation of the modes.  Reference~\cite{flebus2020non,gunnink2022nonlinear}  showed that magnonic topological edge states yield clear experimental signatures in the nonlinear spin dynamics when the bulk spectrum remains real while an edge state acquires nonzero negative imaginary eigenenergy (i.e., signaling a growth of the magnon population). In Ref.~\cite{flebus2020non,gunnink2022nonlinear}  such scenario is realized in a one-dimensional $\mathcal{PT}$-symmetric system whereas the boundary naturally break the symmetry, leading to complex-conjugate edge states coexisting with a purely real bulk spectrum.

In our system, however, open boundaries preserve the wavevector dependent pseudo-Hermiticity~\eqref{273}, and, as a consequence, the bulk and edge states are simultaneously  purely real or complex conjugate paired.

\section{Conclusion and outlooks\label{Sec:conclusion}}

In this work, we investigate the properties of a spin-orbit-coupled ferromagnetic bilayer whose magnetization dynamics is driven via spin current injection. We derive an effective non-Hermitian Hamiltonian by linearizing the coupled Landau-Liftschitz-Gilbert equations of motion while assuming balance of gain and loss between the top and bottom layers. 
We find that, due to fine-tuned balance of gain and loss, the energy levels come always real or complex-conjugate paired. Although the spectrum  hints at  $\mathcal{PT}$-symmetry or (parameter-independent) pseudo-Hermiticity, a more detailed symmetry analysis shows that our system displays a $\mathbf{k}$-dependent pseudo-Hermiticity. 

By tuning the parameter controlling the gain/loss of magnetization dynamics, we observe that a multitude of exceptional points can appear in the first Brillouin zone of the honeycomb lattice. It is well known that  a system displays a strong spectral response to perturbations at an exceptional point.
Thus, the presence of a relevant number of EPs might significantly affect properties routinely probed in spintronics setups whose strength depends on  a Brillouin zone integration of  response functions. Future work should address a Green's function theory for spin transport in the presence of EPs. In particular, it could be intriguing to investigate how thermal spin transport might be affected by coherently driving EPs with, e.g., an external RF field~\cite{liu2019microwave}.

We find that the Dzyaloshinskii-Moriya interaction yields topologically nontrivial non-Hermitian magnon bands. Furthermore, we show  that a wavevector-dependent pseudo-Hermiticity can lead to a purely real bulk and edge spectra, analogously to Hermitian topological insulating phases. However, exact diagonalization for both a nanoribbon with zigzag and armchair boundary conditions shows that the system can not realize  a purely real bulk spectrum with a lasing edge state for a relevant number of momenta, which would allow to observe topological signatures in the nonlinear spin dynamics~\cite{flebus2020non,gunnink2022nonlinear}.
We leave to future work a systematic analysis of the symmetry constrains yielding the above mentioned scenario.

\section{ACKNOWLEDGEMENTS}
The authors thank R. A. Duine and P. M. Gunnink for helpful discussions. 
This work was supported  by the National Science Foundation under Grant No. NSF DMR-2144086.

\bibliographystyle{apsrev4-2}

\end{document}